\documentclass{llncs}

\usepackage[T1]{fontenc}
\usepackage[utf8]{inputenc}
\usepackage{todonotes}
\usepackage{hyperref}

\begin{document}

\title{Bibliometric-enhanced Information Retrieval:\\5th International BIR Workshop}
\titlerunning{BIR 2017}  
%
\author{Philipp Mayr\inst{1} \and Ingo Frommholz\inst{2}\and Guillaume Cabanac\inst{3}}
\authorrunning{Mayr et al.} 
\institute{GESIS - Leibniz-Institute for the Social Sciences, Cologne, Germany,\\
\email{philipp.mayr@gesis.org}
\and
Institute for Research in Applicable Computing, University of Bedfordshire,\\ Luton, UK,\\\email{ifrommholz@acm.org}
\and
University of Toulouse, Computer Science Department, IRIT UMR 5505, France\\
\email{guillaume.cabanac@univ-tlse3.fr}
}


\maketitle              

\begin{abstract}	
Bibliometric-enhanced Information Retrieval (BIR) workshops serve as the annual gathering of IR researchers who address various information-related tasks on scientific corpora and bibliometrics. The workshop features original approaches to search, browse, and discover value-added knowledge from scientific documents and related information networks (e.g., terms, authors, institutions, references). We welcome contributions elaborating on dedicated IR systems, as well as studies revealing original characteristics on how scientific knowledge is created, communicated, and used. In this paper we introduce the BIR workshop series and discuss some selected papers presented at previous BIR workshops. 
\end{abstract}


\section{Introduction}
Following the successful workshops at ECIR 2014\footnote{\url{http://ceur-ws.org/Vol-1143/}}, 2015\footnote{\url{http://ceur-ws.org/Vol-1344/}}, 2016\footnote{\url{http://ceur-ws.org/Vol-1567/}} and JCDL 2016\footnote{\url{http://ceur-ws.org/Vol-1610/}}, respectively, this workshop is the fifth in a series of events that brought together experts of communities which often have been perceived as different ones: bibliometrics~/ scientometrics~/ informetrics on the one hand and information retrieval on the other hand. Our motivation as organizers of the workshop started from the observation that main discourses in both fields are different, that communities are only partly overlapping and from the belief that a knowledge transfer would be profitable for both sides \cite{Mayr2015,Wolfram15}. The need for researchers to keep up-to-date with their respective field given the highly increasing number of publications available has led to the establishment of scientific repositories that allow us to use additional evidence coming for instance from citation graphs to satisfy users' information needs.

The first BIR workshops in 2014 and 2015 set the research agenda by introducing each group to the other, illustrating state-of-the-art methods, reporting on current research problems, and brainstorming about common interests. The third workshop in 2016~\cite{Mayr2016} further elaborated on these themes. For the fourth workshop, co-located with the ACM/IEEE-CS Joint Conference on Digital Libraries (JCDL) 2016, we broadened the workshop scope and interlinked the BIR workshop with the natural language processing (NLP) and computational linguistics field~\cite{Cabanac2016}.
This 5th full-day BIR workshop at ECIR 2017 aims to foster a common ground for the incorporation of bibliometric-enhanced services (incl.\ text mining functionality) into scholarly search engine interfaces. In particular we address specific communities, as well as studies on large, cross-domain collections like Mendeley and ResearchGate. This fifth BIR workshop again addresses explicitly both scholarly and industrial researchers.

\section{Goals, Objectives and Outcomes}
Our workshop aims to engage the IR community with possible links to bibliometrics. Bibliometric techniques are not yet widely used to enhance retrieval processes in digital libraries, yet they offer value-added effects for users~\cite{Mutschke2011}.  Hence, our objective is to bring together information retrieval, information seeking, science modelling, network analysis, and digital libraries to apply insights from bibliometrics, scientometrics, informetrics and text mining to concrete, practical problems of information retrieval and browsing. We discuss some examples from previous workshops in Section~\ref{sec:selected}.
More specifically we ask questions like:
\begin{itemize}
	\item How can we generalize paper tracking on social media?\\a.k.a.\ altmetrics on steroïds: beyond DOI spotting.
    \item How can we detect fake reviews \cite{BartoliEtAl2016} to sustain the peer review process?
    \item How can we improve homonym detection (e.g., \href{http://dblp.uni-trier.de/pers/hd/l/Li:Li}{Li Li}) in bibliographic records~\cite{Momeni2016}?
    \item To what degree can we automate fact-checking~\cite{Baker2015,ZiemannEtAl2016} in academic papers?
    \item How can we support researchers in finding relevant scientific literature, e.g.,\ by integrating ideas from information retrieval, information seeking and searching and bibliometrics~\cite{Abbasi/Frommholz:2014,Mutschke2015}?
 	\item How can we build scholarly information systems that explicitly use bibliometric measures at the user interface (e.g. contextual bibliometric-enhanced features \cite{Carevic2016})?
 	\item How can models of science be interrelated with scholarly, task-oriented searching?
 	\item How can we combine classical IR (with emphasis on recall and weak associations) with more rigid bibliometric recommendations~\cite{Zitt15,Beel2016}?
\item How can we create suitable testbeds (like iSearch corpus)~\cite{LarsenL16}?
\end{itemize}

\section{Format and Structure of the Workshop}
The workshop will start with an inspirational keynote ``Real-World Recommender Systems for Academia: The Pain and Gain in Developing, Operating, and Researching them'' by Joeran Beel (Trinity College Dublin, the School of Computer Science and Statistics) to kick-start thinking and discussion on the workshop topic. This will be followed by paper presentations in a format that we found to be successful at previous BIR workshops: each paper is presented as a 10 minute lightning talk and discussed for 20 minutes in groups among the workshop participants followed by 1-minute pitches from each group on the main issues discussed and lessons learned. The workshop will conclude with a round-robin discussion of how to progress in enhancing IR with bibliometric methods.

\section{Audience}
The audiences of IR and bibliometrics overlap \cite{Mayr2015,Wolfram15}. Traditional IR serves individual information needs, and is~-- consequently~-- embedded in libraries, archives and collections alike. Scientometrics, and with it bibliometric techniques, has a matured serving science policy. 
We therefore will hold a full-day workshop that brings together IR researchers with those interested in bibliometric-enhanced approaches. Our interests include information retrieval, information seeking, science modelling, network analysis, and digital libraries. 
The workshop is closely related to the past BIR workshops at ECIR 2014, 2015, 2016 and strives to feature contributions from core bibliometricians and core IR specialists who already operate at the interface between scientometrics and IR. While the past workshops laid the foundations for further work and also made the benefit of bringing information retrieval and bibliometrics together more explicit, there are still many challenges ahead. One of them is to provide infrastructures and testbeds for the evaluation of retrieval approaches that utilise bibliometrics and scientometrics. To this end, a focus of the proposed workshop and the discussion will be on real experimentations~(including demos) and industrial participation. This line was started in a related workshop at JCDL~(BIRNDL~2016), but with a focus on digital libraries and computational linguistics and not on information retrieval and information seeking and searching.

\section{Selected papers and past Keynotes}
\label{sec:selected}
Past BIR workshops had invited talks of several experts working in the field of bibliometrics and information retrieval.  Last year, Marijn Koolen  gave a  keynote on  ``Bibliometrics in online book discussions: Lessons for complex search tasks''~\cite{Koolen16}. Koolen explored the potential relationships between book search information needs and bibliometric analysis and introduced the Social Book Search Lab, triggering a discussion on the relationship between book search and bibliometric-enhanced IR. In 2015, the keynote ``In Praise of Interdisciplinary Research through Scientometrics''~\cite{Cabanac15} was given by Guillaume Cabanac. Cabanac accentuated the potential of interdisciplinary research at the interface of information retrieval and bibliometrics. He came up with many research questions that lie at the crossroad of scientometrics and other fields, namely information retrieval, digital libraries, psychology and sociology. 

Recent examples of BIR workshop publications have shown the potential of informing the information retrieval process with bibliometrics. These examples comprise topics like IR and recommendation tool development, bibliometric IR evaluation and data sets, and the application and analysis of citation contexts for instance for cluster-based search.

As an example of recommendation tool development utilising bibliometrics, Wesley-Smith et al.~\cite{Wesley-SmithDW15} describe an experimental platform constructed in collaboration with the repository Social Science Research Network~(SSRN) in order to test the effectiveness of different approaches for scholarly article recommendations.
Jack et al.~\cite{JackLHK14} present a case study on how to increase the number of citations to support claims in Wikipedia. They analyse the distribution of more than 9 million citations in Wikipedia and found that more than 400,000 times an explicit marker for a needed citation is present. To overcome this situation they propose different techniques based on Bradfordizing and popularity number of readers in Mendeley to implement a citation recommending system. The authors conclude that a normal keyword-based search engine like Google Scholar is not sufficient to be used to provide citation recommendation for Wikipedia articles and that altmetrics like readership information can improve retrieval and recommendation performance.


Utilising a collection based on PLOS articles, Bertin and Atanassova~\cite{BertinA14} try to further unravel the riddle of meaning of citations. The authors analyse the word use in standard parts of articles, such as Introduction, Methods, Results and Discussion, and reveal interesting distributions of the use of verbs for those sections. The authors propose to use this work in future citation classifiers, which in the long-term might also be implemented in citation-based information retrieval.

As an application of citation analysis, Abbasi and Frommholz~\cite{AbbasiF14} investigate the benefit of combining polyrepresentation with document clustering, where representations are informed by citation analysis. The evaluation of the proposed model on the basis of the iSearch collection shows some potential of the approach to improve retrieval quality. A further application example reported by Nees Jan van Eck and Ludo Waltman~\cite{EckW14} considers the problem of scientific literature search. The authors suggest that citation relations between publications can be a helpful instrument in the systematic retrieval process of scientific literature. They introduce a new software tool called CitNetExplorer that can be used for citation-based scientific literature retrieval. To demonstrate the use of CitNetExplorer, they employ the tool to identify publications dealing with the topic of ``community detection in networks''. They argue that their approach can be especially helpful in situations in which one needs a comprehensive overview of the literature on a certain research topic, for instance in the preparation of a review article.

Howard D.\ White proposes an alternative to the well-known bag of words model called \emph{bag of works}~\cite{White16}. This model can in particular be used for finding similar documents to a given seed one. In the bag of works model, tf and idf measures are re-defined based on (co-)citation counts. The properties of the retrieved documents are discussed and an example is provided.

\section{Output}
In 2015 we published a first special issue on ``Combining Bibliometrics and Information Retrieval'' in Scientometrics~\cite{Mayr2015}. A special issue on ``Bibliometrics, Information Retrieval and Natural Language Processing in Digital Libraries'' is currently under preparation for the International Journal on Digital Libraries. For this year's ECIR workshop we continue the tradition of producing follow-up special issues. Authors of accepted papers at this year's BIR workshop will again be invited to submit extended versions to a special issue on ``Bibliometric-enhanced IR'' to be published in Scientometrics.

\bibliographystyle{splncs}
\bibliography{bibdb}
\end{document}